\documentclass[useAMS]{mn2e}
\usepackage{graphicx}
\usepackage{txfonts}
\usepackage{ctable}
\usepackage{threeparttable}

\title[The evolution of high-z QSOs]{High-redshift quasars host galaxies:
  is there a stellar mass crisis?}
\author[Valiante et al.]{Rosa Valiante$^{1}$\thanks{E-mail:
rosa.valiante@oa-roma.inaf.it}, Raffaella Schneider$^{1}$, Stefania Salvadori$^{2}$,
Simona Gallerani$^{3}$\\
$^{1}$INAF - Osservatorio Astronomico di Roma, via di Frascati 33, 00040, Monteporzio Catone, 
Italy\\
$^{2}$Kapteyn Astronomical Institute, University of Groningen, Landleven 12, 9747 AD Groningen, The Netherlands\\
$^{3}$Scuola Normale Superiore di Pisa, Piazza dei Cavalieri 7, 56126 Pisa, Italy \\
}

\begin{document}

\date{Accepted . Received }

\pagerange{\pageref{firstpage}--\pageref{lastpage}} 
\pubyear{2014}

\maketitle

\begin{abstract}
We investigate the evolutionary properties of a sample of quasars at
$5<z<6.4$ using the semi-analytical hierarchical model 
\textsc{GAMETE/QSOdust}.
We find that the observed properties of these quasars are well reproduced 
by a common formation scenario in which stars form according to a standard 
IMF, via quiescent star formation and efficient merger-driven bursts, while 
the central BH grows via gas accretion and BH-BH mergers.
Eventually, a strong AGN driven wind starts to
clear up the ISM of dust and gas, damping the star formation and 
un-obscuring the line of sight toward the QSO.
In this scenario, all the QSOs hosts have final stellar masses in the range 
$(4-6)\times 10^{11}$ M$_{\odot}$, a factor 3-30 larger than the upper limits 
allowed by the observations. 
We discuss alternative scenarios to alleviate this apparent tension: the most 
likely explanation resides in the large uncertainties that still affect 
dynamical mass measurements in these high-z galaxies. 
In addition, during the transition between the starburst-dominated 
and the active QSO phase, we predict that $\sim 40\%$ of the progenitor 
galaxies can be classified as Sub Millimeter Galaxies, although their number 
rapidly decreases with redshift.  
\end{abstract}

\begin{keywords}
Galaxies: evolution, high-redshift, ISM; quasars: general; 
ISM: dust, extinction; submillimetre: galaxies 
\end{keywords}

\section{Introduction}
High redshift quasars (QSOs), the bright active nuclei (AGN) of galaxies, 
are among the most important sources of information on the Universe at early 
cosmic epochs. 
Their large luminosities ($>10^{47}$ erg/s) at $\rm z>6$ imply 
that super massive black holes (SMBHs), with masses $\rm >10^9 M_\odot$ 
(e.g. Fan et al. 2001, 2003; Willott et al. 2007), were already in place; 
so far, the most distant QSO observed is ULAS J1120+0641 at 
$z=7.085$ (Mortlock et al. 2011) and its estimated BH mass, 
$\rm \sim 2\times 10^9 M_\odot$, must have formed in less than $\sim 800$ Myr. 

In the Local Universe, tight correlations between BH masses and several 
properties of their host galaxies (such as the stellar bulge mass, luminosity, 
and velocity dispersion) have been observed, suggesting a common formation 
scenario (co-evolution) for galaxies and BHs at their center, but the physical 
drivers of this co-evolution may be different in different kind of galaxies 
(for a recent thorough review see Kormendy \& Ho 2013). 

The mean BH-stellar bulge mass ratio, $\rm M_{BH}/M_{star}$,
estimated from observations of a sample of $\rm z>5$ quasar is about 10 times 
higher than the local value (Wang et al. 2010). 
Although these results could be biased by large uncertainties and observational 
selection effects (Lauer et al. 2007; Volonteri \& Stark 2011), a 
possible explanation to this discrepancy 
could be that BHs form before or faster than their host 
galaxy stellar bulges at these early epochs (Lamastra et al. 2010).

Given the large energy released by BH accretion, QSOs activity is 
expected to play a critical role in shaping the star formation history (SFH) 
of the host galaxy, ultimately self-regulating BH growth.
Indeed, massive, large scale gas outflows associated to QSOs have been 
recently detected from the Local Universe up to $\rm z\sim 6.4$ 
(Feruglio et al. 2010; Nesvadba et al. 2010, 2011; 
Maiolino et al. 2012; Cicone et al. 2012).

Finally, observations of high-z galaxies and quasars reveal a rapidly enriched 
interstellar medium (ISM). 
Emission line ratios are seen to trace super-solar gas metallicities (up to
$\rm \sim 10 \, Z_\odot$) in Broad (Nagao et al. 2006, 2012; Juarez et al. 2009)
and Narrow (Matsuoka et al. 2009) Line Regions, with almost no redshift 
evolution. Dust thermal emission for a sample of $\rm 5<z<7.1$ quasars in the 
Sloan Digital Sky (SDSS) survey has been detected through millimetre (mm) and 
sub-mm observations, suggesting large masses of dust, of a few $\rm 10^8 
M_\odot$ 
(e.g. Bertoldi et al. 2003; Priddey et al. 2003; Robson et al. 2004; 
Beelen et al. 2006; Wang et al. 2008). 
Moreover, significant dust reddening is present along several high-z AGN and 
Gamma Ray Bursts (GRBs) lines of sight (Gallerani et al. 2010; Stratta et al.
2011; Hjort et al. 2013)

From the theoretical point of view, many efforts have been made so far to
shed light on the formation and growth of the first galaxies, their central SMBHs 
and/or the evolution of the ISM properties. 
Some of these issues have been approached with 
semi-analytical models, providing insights either on the evolution of the 
dust and metals in the host galaxies (e.g. Morgan \& Edmunds 2003; Hirashita 
\& Ferrara 2002; Dwek, Galliano \& Jones 2007; Valiante et al. 2009;  Gall et al. 2010, 2011; 
Dwek \& Cherchneff 2011; Mattsson 2011, Pipino et al 2011, Calura et al. 2013) or 
on the formation and growth of the central BHs (e.g. Somerville et al. 2008; 
Menci et al. 2008; Lamastra et al. 2010, Volonteri et al. 2003; Volonteri \& 
Rees 2006; Devecchi et al. 2010, 2012). 
On the other hand, high-resolution and/or zoomed-in numerical simulations 
represent the most sophisticated tools to 
investigate the physical processes which drive and regulate the BH-galaxy 
co-evolution (e.g. Di Matteo et al. 2005, 2008, 2012; Hopkins et al. 2006, 
2011; Booth \& Shaye 2009; Li et al 2007, 2008; Sijacki et al 2009; 
DeBhur et al. 2010, DeGraf et al. 2012; Bellovary et al. 2011, 2013). 
These studies have demonstrated that in order to investigate
the formation and growth of black holes and their host galaxies, 
one must embed their evolution in a cosmological setting (Volonteri \& Bellovary 2013).

To interpret the observed/inferred properties of high-z quasars mentioned above, 
all the aspects of BH-galaxy co-evolution must be explained and modelled 
simultaneously in an adequate, self-consistent cosmologically evolving 
context. 
As an attempt to progress in this direction, we have developed a 
semi-analytical model for the formation and evolution of high redshift quasars,
\textsc{GAMETE/QSOdust} (Valiante et al. 2011),  
which is able to follow the hierarchical assembly and merger history
of both the host galaxy and its central BH. The formation and evolution
of the chemical properties of the host galaxy, namely the mass of gas, stars, 
metals and dust are consistently followed in the model.
  
As a pilot study we have 
used this model to investigate possible evolutionary scenarios of one of
the best studied quasars, SDSS J1148+5251 (hereafter J1148), observed at 
redshift $z=6.4$. In Valiante et al. (2011) we pointed out that the 
the observed properties of J1148, such as the BH mass, the mass of gas and 
dust in the ISM, are reproduced only if distinct evolutionary paths are 
followed. 
In particular, if stars form according to a Larson Initial Mass Function (IMF), 
with a characteristic mass $\rm m_{ch}=0.35 \, M_\odot$ (see Eq~\ref{eq:imf}), 
that we call \textit{standard IMF }, at the end of its evolution, 
J1148 settles on the local $\rm M_{BH} - M_{star}$ relation and the stellar 
mass at $\rm z = 6.4$ exceeds the upper limit set by the observed dynamical 
mass by a factor of 3-10 (model B3). 
Alternatively,  a Larson IMF but with a larger characteristic mass of 
($\rm m_{ch}=5 \, M_\odot$), that we call \textit{top-heavy IMF}, can be 
assumed (model B1). 
Note that these results have been recently confirmed by Calura et al. (2013).

Although a top-heavy IMF (model B1) enables to reproduce the observed 
dust mass and the deviation of J1148 from the local Mbh-Mstar relation, the 
predicted star formation rate at $z = 6.4$ is $< 100 $ M$_\odot/$yr, more than 
one order of magnitude smaller than the observed value 
($\sim 3000$ M$_\odot/$yr). 
This rate is too small to power the observed Far Infra-Red (FIR)  luminosity 
(Schneider et al. 2014b). 

Hence, our studies suggest that J1148
has built up a large stellar mass of $\sim 4\times 10^{11}$ M$_\odot$ 
over its evolution and that its final $\rm M_{BH}/M_{star}$ is $\approx 0.007$.
However, this conclusion implies that current
dynamical mass measurements may have missed an
important fraction of the host galaxy stellar mass.

In the present work, we extend our analysis to a larger sample of quasars observed 
between redshift $\rm z=5$ and $\rm z=6.4$. The main aim is to 
assess the robustness of our previous findings and to test if the best-fit scenario 
proposed to explain the formation and chemical evolution of J1148 and its host galaxy 
(Valiante et al. 2011; Schneider et al. 2014b) can  explain the properties of the first 
QSOs in general.

The paper is organized as follows: in section 2 we describe the sample of high-redshift quasars 
that we have selected; a brief presentation of the model, with the new features that we have recently
implemented, is given in section 3; the results of the analysis are presented in section 4 and 5 and
discussed in section 6, where we also draw our conclusions.

In this analysis, we assume a Lambda Cold Dark Matter ($\Lambda$CDM) 
cosmology  with $\Omega_m = 0.24$, $\Omega_\Lambda = 0.76$, $\Omega_b = 0.04$, 
and $H_0 = 73$ km s$^{-1}$ Mpc$^-{1}$. 

\section{The sample of high redshift quasars}\label{sec:sample}

\begin{table*}
  \caption{Inferred physical properties of the sample of 12 QSOs collected 
   from the literature. References for each object are given in the text. 
   Objects highlighted in bold face are the ones that we have selected for the 
   analysis done in Section \ref{sec:results}.}\label{tab:sample}
  \begin{tabular} {|c|c|c|c|c|c|c|c|} \hline
    QSO name & z & $\rm M_{BH}$ ($10^9$ M$_\odot$) & $\rm M_{\rm H2}$($10^{10}$M$_\odot$) & $\rm M_{dyn} sin^2i$ ($10^{10}$M$_\odot$) & L$_{\rm FIR}$ (10$^{13}$L$_\odot$) & SFR (10$^3$ M$_\odot/yr$) &$\rm M_{\rm dust}$ ($10^{8}$M$_\odot$)  \\ \hline
    \textbf{SDSS J0338+0012} & 5.0  & $2.5\pm 0.7$ & 2.5$\pm$1.8 & 8.2$\pm$4.1 & 1.2$\pm$0.2 &1.06$\pm$0.23 & 6.8$^{+4.4}_{-3.7}$ \\
    SDSS J0129-0035 & 5.77  & 0.17$\pm$ 0.051 & 1.24$\pm$1.01 & 2.6$\pm$1.6 & 1.55$\pm$0.3 & 1.4$\pm$0.3 & 2.4$^{+1.2}_{-1.18}$ \\
    \textbf{SDSS J0927+2001} & 5.79  & 2.3$\pm$ 0.69 & 4.8$\pm$3.9 & 11.4$\pm$5.0 & 1.67$\pm$0.3 & 1.5$\pm$0.4 & 6.9$^{+3.7}_{-3.4}$  \\ 
    \textbf{SDSS J1044-0125} & 5.78  & $7.5^{+3.0}_{-2.0}$ & 0.89$\pm$0.7 & 0.84$\pm$0.5 & 1.17$\pm$0.3 & 1.08$\pm$0.24& 1.8$^{+0.9}_{-0.9}$ \\ 
    SDSS J0840+5624 & 5.84  & 1.98$\pm$ 0.59 & 1.5$\pm$1.4 & 24.3$\pm$10.7 & 2.08$\pm$0.4 & 1.9$\pm$0.4& 3.2$^{+1.6}_{-1.6}$  \\ 
    SDSS J1425+3254 & 5.89  & 1.2$\pm$ 0.36 & 2.0$\pm$1.6 & 15.3$\pm$8.2 & 1.44$\pm$0.3 & 1.3 $\pm$0.3 & 2.2$^{+1.08}_{-1.09}$ \\ 
    SDSS J1335+3533 & 5.9  & 2.35$\pm$ 0.71 & 1.86$\pm$1.4 & 3.16$\pm$1.02 & 1.52$\pm$0.3 & 1.4$\pm$0.3 & 2.4$^{+1.15}_{-1.20}$ \\
    \textbf{SDSS J2310+1855} & 6.0  & $3.4\pm 0.6$ & 5.4$\pm$3.8 & 6.8$\pm$1.9 & 4.0$\pm$0.6 & 3.7$\pm$0.3 & 9.0$^{+9.4}_{-4.4}$  \\ 
    SDSS J2054-0005 & 6.04  & 1.2$\pm$ 0.36 & 1.23$\pm$1.02 & 4.3$\pm$2.6 & 1.51$\pm$0.34 & 1.4$\pm$0.3 & 2.35$^{+1.12}_{-1.14}$ \\ 
    ULAS J1319+0950 & 6.13  & 2.1$\pm$ 0.63 & 1.6$\pm$1.3 & 9.5$\pm$4.3 & 2.6$\pm$0.4 & 2.4$\pm$0.5 & 4.05$^{+1.92}_{-1.96}$ \\ 
    SDSS J1048+4637 & 6.23  & $3.0\pm 0.9$ & 1.1$\pm$0.9 & 4.5$\pm$3.2 &  1.82$\pm$0.25 & 1.7$\pm$0.4 & 2.8$^{+1.32}_{-1.36}$ \\
    \textbf{SDSS J1148+5251} & 6.42  & $3.0^{+3.0}_{-1.0}$ & 2.3$\pm$1.9 & 3.4$\pm$1.3 & 2.2$\pm$0.33 & 2.0$\pm$0.5 & 3.4$^{+1.38}_{-1.54}$ \\ \hline
  \end{tabular}
\end{table*}

We have collected from the literature a sample of 12 quasars 
observed between $z=5$ and $z=6.4$ whose BH mass is known and which are 
detected through CO and dust continuum emission. 
Their main physical properties are summarized in Table~1.

\subsection{Black hole mass}
\label{sec:BHmass}

The masses of the BHs given in Table \ref{tab:sample} are taken from the 
literature. 
For QSOs J0338 and J1148, M$_{\rm BH}$ is computed with 
virial estimators using MgII and CIV line widths.
Different virial estimators or scaling relations may provide a factor
of $\sim 2-3$ difference in the estimated BH mass (Barth et al. 2003;
Dietrich \& Haman 2004; De Rosa et al. 2011). 
This is the case of J1148, for which a mass of $(2-3)\times 10^9$ 
M$_\odot$ is obtained from MgII-based scaling relations (Willott et al. 2003; 
Barth et al. 2003) while a larger value ($6\times 10^9$ M$_\odot$) is obtained 
using the CIV line width (Barth et al. 2003). 
For J0338, different emission lines and/or scaling relations instead provide 
similar results: $(2.3-2.7)\times 10^9$ M$_\odot$ and $2.5\times 10^9$ M$_\odot$
are estimated using two different CIV-based scaling relations (Dietrich \& 
Hamann 2004) and the MgII line emission (Maiolino et al. 2007).   
The BH masses of J1148 and J0338 quoted in Table \ref{tab:sample} are the 
average of the different values given in the literature.\\ 
The BH mass of J1044 has been computed both from the bolometric 
luminosity, L$_{\rm bol}$ ($5.6\times 10^9 M_\odot$, Priddey et al. 2003; 
$6.4\times 10^9 M_\odot$, Wang et al. 2010) and from the quasar CIV line 
emission ($10.5\times 10^9 M_\odot$, Jiang et al. 2007) and we adopted the 
average value.
The BH mass of the most luminous quasar of the sample, J2310, is the 
average of two different values from Wang et al. 2013 ($2.8\times 10^9 M_\odot$)
and Fan et al. in prep. ($4\times 10^9 M_\odot$, private communication).
Finally, for the remaining quasars in the
sample, we report the values inferred by Wang et al. (2008, 2010, 2013) from 
L$_{\rm bol}$ assuming Eddington-limited accretion. 
No errors are given in the literature for these objects. 
Since typical uncertainties on the BH masses quoted above are about 
$(20-40)\%$, we assume for these quasars a $30\%$ error.

\subsection{Molecular gas and dynamical constraints}
\label{sec:COobs}

The molecular gas mass, $\rm M_{H_2}$, in galaxies is usually inferred from 
the CO($\rm J=1-0$) emission line luminosity, through the relation 
$\rm M_{H_2}= \rm \alpha_{CO} \times L'_{\rm CO(1-0)}$ where $\rm \alpha_{CO}$ is 
the CO luminosity to $\rm H_2$ mass conversion factor 
(see e.g. Solomon \& Vanden Bout 2005 and Carilli \& Walter 2013). 
For those sources in which only $\rm J>1$ line luminosities are available, we 
convert the lowest CO($\rm J-(J-1)$) transition detected to the 
CO($\rm J=1-0$) luminosity using the CO excitation ladder (lines ratios) 
observed in high redshift quasars, including J1148 (Riechers et al. 2009; 
Wang et al. 2010; Carilli \& Walter 2013)\footnote{
We adopt 
$L'_{\rm CO(2-1)}/L'_{\rm CO(1-0)}=0.99$ for QSOs J0840 and J0927 and 
$L'_{\rm CO(3-2)}/L'_{\rm CO(1-0)}=0.97$ for J1148 and J1048 (Carilli \& Walter 
2013). For higher transitions we assume $L'_{\rm CO(5-4)}/L'_{\rm CO(1-0)}=0.88$ 
in the case of J0338 and $L'_{\rm CO(6-5)}/L'_{\rm CO(1-0)}=0.78$ for the 
remaining objects (Riechers et al. 2009).}.
However, this represents only a minor correction in estimates of the 
molecular gas mass in quasar hosts (e.g. Riechers et al. 2009, 2011; 
Wang et al. 2010). 
The major source of uncertainty  is represented by the unknown  
$\alpha_{\rm CO}$ value.
This factor strongly depends on galaxy properties, such as the gas 
metallicity, temperature, excitation, and velocity dispersion (see the review 
by Bolatto et al. 2013), and direct measurements of its value are currently 
unavailable at high redshifts.

In this work, we adopt the  
value $\alpha_{\rm CO}=0.8\pm 0.5$ M$_\odot/$(K km s$^{-1}$ pc$^2$) which has 
been suggested to trace molecular gas in Ultra Luminous Infra Red Galaxies 
(ULIRGs; Solomon et al. 1997; Downes \& Solomon 1998).
A ULIRGs-like conversion factor is usually assumed as a good approximation 
for high redshift galaxies, including Sub Millimetre Galaxies (SMGs) and QSOs 
(Tacconi et al. 2008; Bothwell et al. 2010; Ivison et al. 2011; Magdis et 
al. 2011; Magnelli et al. 2012). 
Errorbars in Table \ref{tab:sample} account for the uncertainty on the 
observed CO emission line flux ($10-30\%$) and for the large scatter 
($>60\%$) in the adopted $\alpha_{\rm CO}$ conversion factor.\\

The dynamical mass has been estimated from CO observations, using the 
formula $\rm M_{dyn} = R\, v^2_{\rm cir}/G$ (Neri et al. 2003; Walter et al. 
2004; Solomon \& Vanden Bout 2005) where $\rm R$ and $\rm v_{circ}$ are the 
molecular disk radius and its maximum circular velocity, respectively.
We take $\rm v_{\rm circ}=(3/4)$ FWHM$_{\rm CO}/$sin $i$ (Wang et al. 2010) 
where FWHM$_{\rm CO}$ is the Full Width at Half Maximum of the CO lines 
and $i$ is the disk inclination angle.
For QSO J1148, the CO emission is spatially resolved out to a radius of about 
$\rm R=2.5$ kpc from the central source and the emitting gas is assumed to 
form an inclined disk with $i=65^o$, where $i=0$ indicates face on disks 
(Walter et al. 2004). 
Since J1148 is the only objects for which the disk radius can be constrained
by the observations, we adopt $\rm R=2.5$ kpc for all the other quasars
in the sample.
Differences in the values of $M_{\rm dyn}$ sin$^2 i$ quoted in Table 
\ref{tab:sample} with respect to previous estimates given in the literature 
are primarily due to different ways of estimating $v_{circ}$ from the CO FWHM 
and/or different assumptions on $\rm R$ (Maiolino et al. 2007; Walter et al. 2004). 
We will discuss the dependence on the assumed inclination angle below.

The stellar mass (dynamical bulge) of the quasar hosts is computed as 
$\rm M_{star} = M_{dyn}- M_{H_2}$.
Strictly speaking, this should be considered as an upper limit to the stellar 
mass, both because we are neglecting the contributions of atomic gas and dark 
matter (see the discussion in Section 6), and we are using a conversion factor 
which maximizes M$_{\rm star}$. In fact, had we used the Milky Way value 
$\alpha_{\rm CO}=4.3$ M$_\odot$ (K km s$^{-1}$ pc$^2$)$^{-1}$, 
the inferred M$_{\rm H_2}$ would be higher, and consequently M$_{\rm star}$ lower 
than the values adopted here.
An average uncertainty of $\sim 50\%$ is assigned to both the 
inclination-corrected dynamical mass M$_{\rm dyn}$ and stellar mass (Walter et 
al. 2004).

In Fig. \ref{fig:bhrel} we show the $\rm M_{BH}-M_{dyn}$ relation of the selected 
high-redshift quasars compared with data and empirical fit obtained for local 
galaxies by Sani et al. (2011). 
In the upper panel, an inclination angle $i=65^o$ is adopted for all quasars, 
while in the lower panel an average value
of $40^o$ is assigned to all QSOs but J1148 (Wang et al. 2010).

As discussed by Wang et al. (2010), observations seem to indicate a deviation 
from the BH-host scaling relations inferred for local galaxies: 
$\rm M_{BH}/M_{dyn}$, $\rm M_{BH}/M_{star}$ and $\rm M_{BH}/\sigma$ ratios are 
about one order of magnitude higher in $\rm 5<z<7$ QSOs, suggesting a faster 
evolution of the first SMBHs with respect to their host stellar bulge (Walter 
et al. 2004; Peng et al. 2006; Riechers et al. 2008; Merloni et al. 2010; 
Wang et al. 2010, 2013). 
However, this result depends on the adopted disk inclination angle:
a lower disk inclination angle results in a shift towards higher dynamical 
masses (and hence stellar masses), but the displacement on the $\rm M_{BH}-M_{dyn}$
or $\rm M_{BH}-M_{star}$ plane still persists for some of of the selected quasars. 
For these quasars, a very small inclination angle, $i<15^o$, a more extended 
disk (larger radius, $\rm R$) or a different, more complex, description of the 
disk geometry would be required to reconcile the inferred dynamical/stellar 
mass with present-day values (Wang et al. 2010, 2013; Valiante et al. 2011).

In addition, it has been pointed out that the observed offset between 
the high redshift QSOs and BH-bulge relation today may be strongly biased by selection effects
(Lauer et al. 2007; Volonteri \& Stark 2011): 
observations at high redshifts are often limited to the most luminous quasars,
actually tracing a narrow range of BH masses and host galaxy properties and thus, not representing 
the whole population of high redshift BHs and hosts.
On the other hand, selecting galaxies of similar luminosity, or BH mass, in 
the local Universe would predict a very similar offset, given the scatter in 
the BH-host correlations.

\subsection{Dust mass, FIR luminosity and SFR}
\label{sec:FIRobs}

The mass of dust, $\rm M_{dust}$, of each quasar in the sample is estimated 
from the rest-frame FIR flux density, assuming optically thin emission:
\begin{equation}
M_{\rm dust} = \frac{S_{\nu_0} \, d_{\rm L}^2(z)}{(1+z) \, \kappa_{\rm d} (\nu) \, B(\nu, T_{\rm d})},
\label{eq:mdust}
\end{equation}
\noindent
where $S_{\nu_0}$ is the flux observed in a given band, $\kappa_{\rm d}(\nu)$ is 
the opacity coefficient per unit dust mass, $B(\nu, T_{\rm d})$ is the Planck 
function for a dust temperature $T_{\rm d}$, and $d_{\rm L}$ is the luminosity 
distance to the source. 
In the Rayleigh-Jeans part of the spectrum, dust radiates as a ’grey-body’ with
$\kappa_{\rm d}(\nu) = \kappa_0 (\nu/\nu_0)^\beta$.
The same $\chi^2$ fit of the spectrum adopted for J1148 in Valiante et al. 
(2011) is used here to derive dust temperature $T_{\rm d}$, $M_{\rm dust}$ and 
FIR luminosity $L_{\rm FIR}$ for quasars detected at different wavelengths in 
the range (350-1200) $\mu \rm m$: J0338, J0927, J2310 and J1148 (Leipski et 
al. 2013; Fan et al. in prep. private communication). 
We adopt different absorption coefficients per unit dust mass, $k_0$ and 
spectral index $\beta$ given in the literature (see Table 1 in Valiante et al. 
2011). 
For these quasars the computed dust temperature ranges between 37 K and 59 K 
and the values quoted in the table are the average dust mass obtained with 
this method.
The other quasars in the sample do not have detailed spectral Energy 
Distrubutions (SEDs) as the previous ones. 
For these objects, we have computed $\rm M_{dust}$ from the continuum 
detection at 1200 $\mu$m, using the same set of parameters, $k_0$, $\beta$, 
and $\rm T_{d}$ obtained from the best-fit of the FIR emission of J1148. 
The resulting mass of dust ranges between $\sim 10^8$ M$_\odot$ and $\sim 10^9$ 
M$_\odot$, with errorbars accounting for the minimum and maximum values.

The SFRs are usually inferred adopting the Kennicutt (1998) relation
between the rate of star formation and the FIR luminosity: 
$L_{\rm FIR}/L_\odot = 5.8 \times 10^9 \, \rm SFR/(M_\odot/yr)$.
This scaling relation assumes that stars have solar metallicity and are 
formed in a 10-100 Myr burst according to a Salpeter IMF. 
For the same FIR luminosity, a factor 2$-$5 lower SFRs are obtained if a 
standard ($m_{ch}=0.35$ M$_\odot$) or a top-heavy ($m_{ch} = 5$ M$_\odot$) IMFs 
are adopted\footnote{Correcting for the different IMF we found FIR-to-SFR 
conversion factors of $10.84\times 10^9$ and $20.86\times 10^9$ for a Larson 
standard and top-heavy IMF, respectively.}.
The SFRs listed in Table~\ref{tab:sample} are obtained from the FIR 
luminosity in the wavelength range $[8-1000] \mu m$, adopting the conversion 
factor of $L_{\rm FIR}/L_\odot = 10.84 \times 10^9 \, \rm SFR/(M_\odot/yr)$ 
required for a Larson IMF with characteristic mass $m_{ch}=0.35$ M$_\odot$. 

In addition, the above relation between $L_{\rm FIR}$ and SFR relies on the 
assumption of starburst-dominated dust heating, when all the FIR luminosity is 
re-emitted by dust heated by young stars.
For this reason, we consider these values as upper limits to the real rates 
of star formation as in luminous QSOs a non negligible contribution to dust 
heating ($30\%$ up to $60\%$) may come from the AGN itself (Wang et al. 2010;
Schneider et al. 2014b), further lowering the estimated SFRs by a factor 
$1.4-2.5$.
 
Finally, as a reference for the ISM gas metallicity, we adopt the value 
inferred from observations of the Narrow Line Regions (NLRs) of high-redshift 
quasars, $\rm Z_{NLR}=1.32^{+0.25}_{-0.22} Z_\odot$ which do not show a 
significant redshift evolution up to $\rm z \approx  6$ (Nagao et al. 2006; 
Matsuoka et al. 2009).
\begin{figure}
\includegraphics [width=9.0cm]{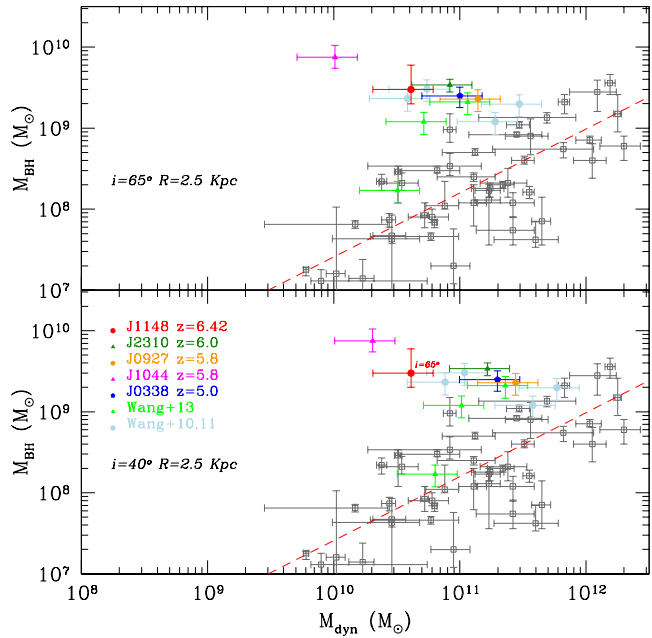}
\caption{Black hole mass as a function of the dynamical mass of the host galaxy. 
         For all quasars, the dynamical mass represents the mass enclosed within 
         a radius of $\rm R=2.5$ kpc. 
         In both panels the solid coloured circles and triangles are data 
         inferred for high redshift quasars from Wang et al. 2010, 2011 and from 
         Wang et al. 2013, respectively (see text and Table \ref{tab:sample} for 
         details). 
         Grey open squares are data for local galaxies with the empirical fit 
         (red dashed line) given by Sani et al. (2011). 
         In the upper (lower) panel the dynamical 
         mass is corrected for a inclination angle $i=65^o$ ($i=40^o$), except
         for J1148, for which we always adopt $i=65^o$.} 
\label{fig:bhrel} 
\end{figure} 

\begin{figure}
\includegraphics [width=9.0cm]{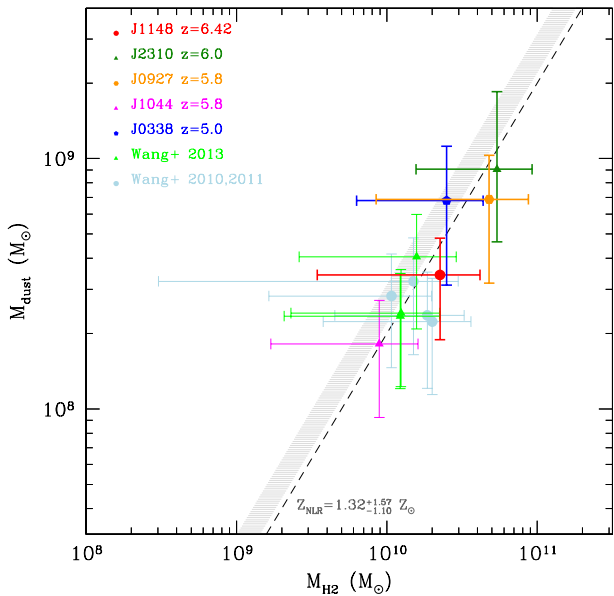}
\caption{Dust mass as a function of the molecular gas mass for high redshift 
         quasars. Data points are the same objects as in Fig. \ref{fig:bhrel}.
         The grey shaded area represents the mass of metals 
         obtained assuming the metallicity of the NLRs 
         $Z_{\rm NLR}=1.32^{+1.57}_{-1.10} Z_\odot$ (Matsuoka et al. 2009) with the 
         dashed line indicating the value obtained for $Z_{\rm NLR}=Z_\odot$.} 
\label{fig:dustMh2} 
\end{figure}

Fig. \ref{fig:dustMh2} shows the dust mass as a function of the H$_2$ mass for
the quasars sample. The data clearly point to a tight correlation between the masses of 
dust and H$_2$, suggesting that dust is mostly associated to the 
dense molecular component of the ISM.
A similar conclusion is drawn by the analysis of the SED of J1148, which
requires that $>50\%$ of dust is distributed into dense clumps (Schneider et al. 2014b).
The grey region in Fig. \ref{fig:dustMh2} represents the mass of gas-phase 
metals obtained as $M_{\rm met} \approx Z_{\rm NLR}\times M_{\rm H2}$, with the dashed 
line indicating the solar abundances, $Z_{\rm NLR}=Z_\odot$. 
All the data points lie close to this region, indicating that the 
large dust masses require almost all the heavy 
elements present in dense clouds traced by CO emission to be in the form of dust 
grains.  

\begin{figure}
\includegraphics [width=8.5cm]{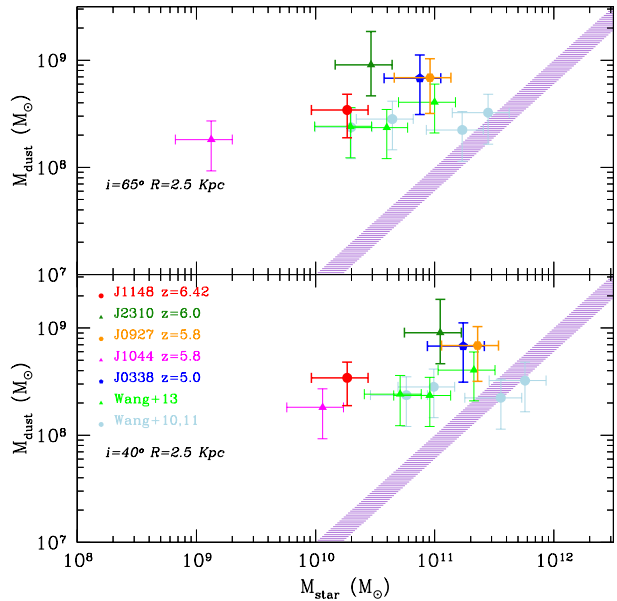}
\caption{Dust mass as a function of the stellar (dynamical bulge) mass. 
         Observational data points are the same as in Fig. \ref{fig:bhrel}.
         The stellar mass is computed as the 
         difference between the molecular mass and the dynamical mass, adopting 
         $i=65^o$ (upper panel) and $40^o$ (lower panel).
         The thick lines represent the maximum dust masses obtained 
         from stellar sources (see text).} 
\label{fig:dustMstar} 
\end{figure}

Fig. \ref{fig:dustMstar} shows the dust mass as a function of the 
stellar mass for the 12 QSOs in the sample. Since stellar masses have been computed as the 
difference between the dynamical and molecular gas masses, their values 
depend on the assumed inclination angle ($i=65^o$ upper panel, or $i=40^o$ 
lower panels). 
The pink shaded area in the figure represents the maximum mass of dust produced 
by a stellar population formed in a single instantaneous burst at solar 
metallicity. 
It is obtained assuming a maximum IMF-averaged dust yield\footnote{The 
IMF-weighted stellar yield is defined as the total mass of dust and metals 
produced per unit stellar mass formed in an instantaneous burst of star 
formation. Here we use the values obtained for stars formed with $Z = Z_\odot$ 
and a Larson IMF with two different values of the characteristic mass, 
$\rm m_{ch} = 0.35, 5 M_{\odot}$ (standard and top-heavy).} 
of $7 \times 10^{-4}  - 10^{-3}$. This includes contribution from supernovae 
(SNe, Bianchi \& Schneider 2007) and intermediate mass (AGB, 
Ferrarotti \& Gail 2006) 
stars and the range spans variations in the adopted IMF (standard/top-heavy).
The figure shows that for the majority of the QSOs in the sample, the
mass released by the stars falls short of the observed value. Not surprisingly,
the only two exceptions are objects whose stellar masses 
are large enough to lie within the scatter of the local $\rm M_{BH} - M_{star}$
relation (see Fig.\ref{fig:bhrel}).
Note that here we are assuming that SNe produce
$[0.1 - 0.6]$ M$_\odot$ of dust but that moderate destruction by the reverse 
shock leads to an effective yield of $[10^{-2} - 10^{-1}]$ M$_\odot$ 
(Bianchi \& Schneider 2007), consistent with observations of SN remnants (see 
Fig. 6 of Schneider et al. 2014a). The effect of a higher SN-dust yield will 
be discussed in Section~\ref{sec:disc}.\\
In addition, here we are neglecting the destruction of dust grains by SN 
shocks in the ISM, hence maximizing the contribution from stellar sources. 
Therefore, alternative processes must play a role. 
Dust condensation in quasar winds (Elvis et al. 2002) and grain growth in 
molecular clouds have been proposed as alternative dust sources. 
The first possibility has not been deeply investigated yet, but Pipino et al. (2011) 
suggest that the 
contribution of the quasar winds to dust production should be negligible with 
respect to that of stellar sources on the large galactic scales. 
On the other hand, dust accretion in molecular clouds is already considered the 
primary non-stellar source of dust in the Milky Way and the Large Magellanic
Cloud (Zhukovska et al. 2008; Zhukovska \& Henning 2013; Schneider et al. 2014a) and
it has also been advocated  to solve the so-called
\textit{dust budget crisis} in high redshift galaxies and quasars, where the dust 
produced by stellar sources is not enough to explain the observed dust 
masses (Michalowski et al. 2010; Valiante et al. 2011;  Pipino et al. 2011; 
Rowlands et al. 2014). 

In summary,  it appears that QSOs at redshifts $5 \le z \le 6.4$ show similar 
properties, both for the central engine (the mass of the BH) and for the host 
galaxy (the dynamical, dust and molecular gas masses), pointing to a common 
evolutionary scenario.
 

\section{Summary of the model}
\label{sec:gamete2p}

In this section, we give a brief summary of the hierarchical 
semi-analytical model \textsc{GAMETE/QSOdust}, presenting  
the additional features that have been recently implemented. 
We refer the interested reader to Valiante et al. (2011) for a more 
complete presentation of the model.

\textsc{GAMETE/QSOdust} describes the co-evolution of BHs and 
their host galaxies, following at the same time the metal and dust
enrichment of the interstellar medium.  
The observed properties of the quasars can be used to constrain the 
set of model parameters which define a specific evolutionary scenario, 
such as the efficiency of star formation and the efficiencies of BH accretion and feedback.

We start with a dark matter halo of $10^{13}$ M$_\odot$, which is believed 
to host $z\sim 6$ SMBHs (Fan et al. 2004). 
This massive halo is decomposed into progressively smaller dark matter progenitors,
backward in time, according to the Extended Press-Schechter (EPS) theory, 
using a binary Monte Carlo algorithm with mass accretion (see Salvadori et al. 
2007 and Valiante et al. 2011 for details)\footnote{Note that, the EPS 
formalism underestimates the abundance of massive halos ($M\geq 10^{12}$ 
M$_\odot$) at redshift $z>1$ with respect to N-body simulations. 
However, the discrepancy between the semi-analytical prediction and numerical 
simulations is within a factor of 2, and typically lower than $30\%$ (Lacey \& 
Cole 1994; Somerville et al. 2000). This does not significantly affect the our 
study.}. 
With this method, we produce several hierarchical merger histories along which 
the gradual build up of the nuclear SMBH and the host galaxy proceeds hand in 
hand. 

At each redshift, seed BHs of $10^4h^{-1}$ M$_\odot$ are assigned to progenitors,
corresponding to $>4\sigma$ density fluctuations, at the time they reach the 
threshold mass required to form stars. As a result, BH seeds are planted in 
only a fraction of progenitor halos at $z > 8$ (see Figure 2 in Valiante et 
al. 2011). The results are not sensitive to the adopted BH seed mass provided 
that $M_{seed} > 10^3 h^{-1}$ M$_\odot$. 
This is due to the interplay between Eddington-limited BH accretion and AGN 
feedback processes which efficiently regulate the BH growth.
The value of this mass threshold as well as the dependence 
on the assumed seed BH mass are discussed in Valiante et al. (2011). 

In modelling the evolution and feedback of the BH, we were guided by several
works, among which Springel, Di Matteo \& Hernquist (2005), Di Matteo et al. 
(2005, 2008) and Sijacki et al. (2007).
Accretion onto the BH is assumed to proceed at the Eddington-limited rate, with
a gas accretion rate defined as the minimum between the Eddington value 
and the accretion rate computed using the Bondi-Hoyle-Lyttleton formula 
(BHL, Eqs.~5 and 6 in Valiante et al. 2011).
The BHL accretion rate is proportional to the BH mass and to
the gas density at the so-called Bondi radius (see Eq~6 in Valiante et al. 
2011). 
This density is computed assuming an isothermal profile with a flat core. 
However, due to the lack of spatial resolution and 
proper physical modelling of the gas conditions close to the BH, we introduce a 
parameter, $\alpha$, that represents the efficiency of gas accretion (Springel 
et al. 2005; Di Matteo et al. 2005).\\
The fraction of energy released by the accreting BH that is transferred to 
the host galaxy, is commonly parametrized as (Di Matteo et al. 2005),
$E_{fdbk}=\epsilon_{w,AGN} \epsilon_r \dot{M}_{accr} c^2$, where $\epsilon_r$ is 
the radiative efficiency, $\dot{M}_{accr}$ is the gas accretion rate and 
$\epsilon_{w,AGN}$ is the wind efficiency tuned to reproduce the final host gas 
mass. Indeed, a robust prediction of the model is that the evolution of the 
nuclear BH and of the host galaxy are tightly coupled by quasar feedback in 
the form of strong galaxy-scale winds (Valiante et al. 2011):
$dM_{ej}/dt = 2\epsilon_{w,AGN} \epsilon_r (c/v_e)^2 \dot{M}_{accr}$.
For J1148, the predicted mass outflow rates are in excellent agreement with the 
observations (Valiante et al. 2012).

At each time, the SFR is assumed to be proportional to the available mass 
of gas, $M_{\rm ISM}$, with a total efficiency, 
$\epsilon = \epsilon_{\rm quies}+\epsilon_{\rm burst}$, that is enhanced 
during major mergers: 
\begin{equation}
{\rm SFR} = M_{\rm ISM} (\epsilon_{\rm quies}+\epsilon_{\rm burst})/t_{\rm dyn}(z) ,
\label{eq:sfr}
\end{equation}
\noindent
where $t_{\rm dyn}(z) = R_{\rm vir}/v_{\rm e}$ is the dynamical time, 
$\epsilon_{\rm quies}$ and $\epsilon_{\rm burst}$ are the quiescent and starburst
efficiencies, respectively. The latter efficiency has been parametrized as a 
normalized Gaussian distribution of the mass ratio of the merging halos 
(Valiante et al. 2011). 

Finally, the enrichment of the host galaxy ISM in metals and dust is computed 
in a self-consistent way assuming that both AGB stars and SNe contribute to 
the total metals and dust mass budget, injecting their products into the ISM 
according to the progenitor stars evolutionary timescales.
The life-cycle of dust implemented in \textsc{GAMETE/QSOdust} is regulated by 
both destruction by interstellar shocks and grain growth. 

The semi-analytical model \textsc{GAMETE/QSOdust} has the advantage to enable 
an extensive investigation of the parameter space on relatively short computational 
times. 
The choice of the main free parameters, namely the efficiency of quiescent 
($\epsilon_{quies}$) and bursting ($\epsilon_{burst}$) star formation, BHL accretion 
($\alpha$) and AGN-driven wind $(\epsilon_{w,AGN})$, all concur in shaping the 
predicted SFH, picturing different plausible evolutionary scenarios. 
In all models presented in Valiante et al. (2011), $\alpha$ and 
$\epsilon_{w,AGN}$ have been chosen to reproduce the final SMBH and gas masses 
while the SF efficiencies have been changed to investigate the effect of 
different SFHs on the evolution of the host galaxy properties.\\
In the present work we will focus on one of these models (B3) 
as our reference model and apply it to the selected sub-sample of quasars,
to check whether they follow a common evolutionary scenario, as suggested by
the similar properties of their central engine and host galaxies.

\subsection{The fiducial scenario}
\label{sec:B3}
Our fiducial model is able to reproduce the final dust 
mass for J1148 providing a sustained SFR as inferred from the FIR luminosity 
of this quasar.
In this model, stars form according to a Larson IMF:
\begin{equation}
\phi(m) \propto m^{-(x + 1)} e^{-m_{ch}/m},
\label{eq:imf}
\end{equation}
\noindent
with $x=1.35$ and a characteristic stellar mass $\rm m_{ch}=0.35 \, M_\odot$, 
normalized to 1 in the $[0.1-100]$ M$_\odot$ mass range (standard IMF). 
At each redshift, stars form quiescently out of the available gas in each 
progenitor galaxy with an efficiency $\epsilon_{\rm quies}=0.1$. 
Each time a major merger occurs, a burst of star formation with efficiency 
$\epsilon_{\rm burst}$ is triggered, providing an additional contribution to 
the stellar mass formed; the value of this parameter depends on the merging 
galaxies mass ratio, reaching a maximum of $\epsilon_{\rm burst} \approx 8$ 
(see Table~2).
For J1148 we assumed $\alpha=200$ and $\epsilon_{w,AGN}=5\times 10^-3$. 
However, the choice of the BH accretion and AGN-driven wind 
efficiencies in the present work will be discussed in the next section.

The final stellar mass is predicted to be $\rm M_{star} = 4\times 10^{11} \, M_\odot$.
This is one order of magnitude larger than $\rm M_{dyn} - M_{H_2}$ for J1148, 
displacing the final $\rm M_{BH}/M_{star}$ from the observational data point toward
the local correlation.

\subsection{New features of the model}
\label{sec:new}

The improvement of the code was guided by two important results:  
\textit{(i)} grain growth in dense molecular gas is required to reproduce 
the mass of dust in J1148, and \textit{(ii)} more than $50\%$ 
of the total dust budget must be distributed in dense clumps to reproduce 
the observed SED of J1148 (Schneider et al. 2014b). 
These conclusions are supported by the relations shown in Fig.~\ref{fig:dustMh2} 
and \ref{fig:dustMstar} discussed in Section~\ref{sec:sample}. 
Indeed, efficient grain growth requires both dust and gas-phase metals to be 
in molecular clouds, where the dust is shielded from the destructive effect of interstellar 
shocks. 

For these reasons, we have improved the prescriptions in our semi-analytical
model in order to mimic a two-phase ISM, namely a diffuse environment (warm/hot
atomic gas) in which the expanding ejecta of SN shocks can destroy the dust 
and a cold-dense medium (the total mass of material in molecular 
clouds) in which star formation and grain growth take place. 
Hereafter, we will refer to these two components of the ISM as 
\textit{diffuse gas} and \textit{dense gas or molecular clouds} (MC).
The total ISM mass $\rm M_{ISM}$ is divided in these two different components, 
$\rm M_{ISM}^{\rm diff}$ and $\rm M_{ISM}^{MC}$.

At the time of the virialization of its host dark matter halo, each
galaxy is composed only by diffuse gas, i.e. $\rm M_{ISM}^{MC}(t_{in})=0$ and 
$\rm M_{ISM}^{diff}(t_{in})=M_{ISM}(t_{in})$. 
Then, the time evolution of these two components as well as the evolution of metals and 
dust in the two phases, is followed by solving a network of differential 
equations. 
In details: 
\begin{equation}
  \rm
  \dot{M}_{\rm ISM}^{\rm diff}(t)= -\dot{M}_{\rm cond}(t)+\dot{R}(t)+ 
  \dot{M}_{inf}(t)-\dot{M}_{ej}^{\rm diff}(t)-\dot{M}_{accr}^{\rm diff}(t),
\end{equation}
\noindent
and
\begin{equation}
\rm  
 \dot{M}_{\rm ISM}^{\rm MC}(t)= -\rmn{SFR}(t)+\dot{M}_{\rm cond}(t)- 
  \dot{M}_{ej}^{\rm MC}(t)-\dot{M}_{\rm accr}^{\rm MC}(t).
\end{equation}
\noindent
As soon as a progenitor galaxy reaches the threshold mass for star formation,
MC are assumed to condense out of the diffuse gas and stars form 
in MC at a rate SFR(t). The stellar products (gas, heavy 
elements and dust) as well as the material accreted from the external medium  are 
returned/injected into the diffuse medium at rates $\rm \dot{R}(t)$, $\rm \dot{Y}_{Z}(t)$, 
$\rm \dot{Y}_{d}(t)$ and $\rm \dot{M}_{inf}$ given by Salvadori et al. (2008) and Valiante et al.
(2009). 
Finally, $\dot{M}_{ej}^{\rm diff}$ and $\dot{M}_{ej}^{\rm MC}$ are the rate at which the gas
is ejected out of the diffuse ISM and MCs, respectively. These two terms are parametrised as:
\begin{equation}
\rm
\dot{M}_{ej}^{\rm diff}=X_{cold}(t) \dot{M}_{ej},
\end{equation}
and 
\begin{equation}
\rm
\dot{M}_{ej}^{\rm diff}=(1-X_{cold}(t)) \dot{M}_{ej},
\end{equation}
where $\dot{M}_{ej}$ is the total gas outflow rate due to both SN and AGN feedback
(see Valiante et al. 2011, 2012 for details on these two contributions) and 
$\rm X_{\rm cold}=M_{\rm ISM}^{\rm MC}(t)/M_{ISM}(t)$ is the cold gas mass fraction.

The physical processes driving the formation of cold dense molecular clouds 
(gravity, magnetic fields, turbulence, shocks, radiation) or controlling their 
survival against disruptive events (cloud-cloud collision and/or winds 
from massive stars) are still far from being understood (see Dobbs et al. 2013 
for a thorough review of the current state of the field). 
A detailed description of the formation and evolution of single clouds
is beyond the scope of this work. Here we are interested in the variation of 
the total amount of material that can be found in MC, constrained to reproduce 
the observed mass of molecular gas.
Therefore, the term $\rm \dot{M}_{cond}$ approximates the cycling of gas between the 
diffuse and dense phases, including both cloud formation and dispersion from/into the 
diffuse gas. 

The rates describing the cycling of material between the two ISM phases can be written 
as
$\rm \dot{M}_{cond}(t) = a \, M_{\rm ISM}^{\rm diff}(t)/t_{form}(t) - b \, M_{\rm ISM}^{\rm MC}(t)/t_{des}(t)$ 
where $\rm a$ and $\rm b$ represent the condensation and disruption efficiencies 
and $\rm t_{form}$ and $\rm t_{des}$ are MC formation and destruction timescales, 
respectively.
These two timescales are assumed to be proportional to the dynamical 
time, $t_{dyn}$, so that in each progenitor halo the SFR
and the rate of mass exchange between the two phases are proportional to each 
other.
\begin{equation}
\rm 
      a \times M_{\rm ISM}^{\rm diff}(t)/t_{form} \propto M_{\rm ISM}/t_{dyn} =  A \, 
      \rmn{SFR}(t),
\end{equation} 
\noindent
and 
\begin{equation}
\rm 
      b \times M_{\rm ISM}^{\rm MC}/t_{des} \propto M_{\rm ISM}/t_{dyn} = B \,
      \rmn{SFR}(t).
\end{equation}
\noindent
Hence,  $\rm \dot{M}_{cond} =\alpha_{MC}\, \rmn{SFR}(t)$, where 
$\rm \alpha_{MC} = A-B >1$, to ensure the formation of MC.

The new features of the code and their dependence on the parameters space have 
been extensively tested applying the model to the Milky Way. 
More details of this analysis and a comparison with observational data
will be given in
a forthcoming work (De Bennassuti et al. 2014).
Here we fix the cloud formation (A) and dispersal (B) coefficients to reproduce the observed 
molecular gas mass. In particular, J1148 requires that $\rm A>>B$ and that  
$\rm A \sim \alpha_{MC}=3.08$.

Similarly, the total mass of heavy elements (gas-phase metals and 
dust) in the two phases evolves according to the following equations: 

\begin{eqnarray}
  \rm
  \dot{M}_{Z}^{\rm diff}(t) & = & \rm -Z_{diff}(t) \, \dot{M}_{\rm cond}(t)+\dot{Y}_Z(t) 
  +Z_{vir}(t) \, \dot{M}_{inf}(t) \nonumber \\
  & & \rm -Z_{diff}(t) \, \dot{M}_{ej}^{\rm diff}(t)-Z_{diff}(t) \, \dot{M}_{accr}^{\rm diff}(t),
\end{eqnarray}
\noindent
and 
\begin{eqnarray}
  \rm 
  \dot{M}_{Z}^{\rm MC}(t) & = & \rm -Z_{MC}(t)\, \rmn{SFR}(t)+Z_{diff}(t)\, \dot{M}_{\rm cond}(t) \nonumber \\ 
   & & \rm -Z_{MC}(t)\, \dot{M}_{ej}^{\rm MC}(t)-Z_{MC}(t)\, \dot{M}_{\rm accr}^{\rm MC}(t),
\end{eqnarray}
where $\rm Z_{diff} = M_Z^{\rm diff}(t)/M_{\rm ISM}^{\rm diff}(t)$ and $\rm Z_{MC} = M_Z^{\rm MC}(t)/M_{\rm ISM}^{\rm MC}(t)$ 
are the metallicities of the diffuse and dense gas.

The major improvement in the chemical network is in the equations describing 
the evolution of the dust. 
In the previous work (Valiante et al. 2011), we assumed that at each time a fixed fraction of 
the total dust mass is shielded against destruction by interstellar shocks and 
can experience grain growth; in addition, no dust ejection in SN-driven or 
BH-driven outflows was implemented.   
These assumptions, although oversimplified, allowed us to reproduce the 
observed dust mass of J1148, pointing out that grain growth in MC dominates
the dust mass even at $\rm z\geq 6$.
In this work, we overcome these two limitations, as we can now follow 
consistently the evolution of the two different phases as a function of time. 
In the new chemical network, dust evolution is described as: 

\begin{eqnarray}
\rm 
  \dot{M}_{d}^{\rm diff}(t) & = & \rm -D_{diff}(t)\dot{M}_{\rm cond}(t)+\dot{Y}_d(t)+ 
  D_{vir}(t)\dot{M}_{inf}(t) \nonumber \\
  & & - \rm M_{d}^{diff}(t)/\tau_d-D_{diff}(t)\dot{M}_{ej}^{\rm diff}(t)-D_{diff}(t)\dot{M}_{accr}^{\rm diff}(t),
\end{eqnarray}
\noindent
and 
\begin{eqnarray}
 \rm 
  \dot{M}_{d}^{\rm MC}(t) & = & \rm -D_{MC}(t)\rmn{SFR}(t)+D_{diff}(t)\dot{M}_{\rm cond}(t)+
  M_{d}^{MCs}/\tau_{acc}\nonumber \\ 
  & & \rm -D_{MC}(t)\dot{M}_{ej}^{\rm MC}(t)-D_{MC}(t)\dot{M}_{\rm accr}^{\rm MC}(t),
\end{eqnarray}
\noindent
where $\rm D_{diff}(t)=M_{d}^{\rm diff}(t)/M_{\rm ISM}^{\rm diff}(t)$ and 
$\rm D_{MC}(t)=M_{d}^{\rm MC}(t)/M_{\rm ISM}^{\rm MC}(t)$ are the dust-to-gas ratios in 
the two phases.
The destruction timescale $\rm \tau_d$ is the same as in Valiante et al. (2011, eq.~13).
Note that the time evolution of the total mass of metals and dust are
$\rm \dot{M}_Z(t)=\dot{M}_{Z}^{diff}(t)+\dot{M}_{Z}^{MC}(t)$ and 
$\rm \dot{M}_d(t)=\dot{M}_{d}^{diff}(t)+\dot{M}_{d}^{MC}(t)$, consistently
with Eqs.~17 and 18 in Valiante et al. (2011). In each phase, 
the mass of the gas-phase metals can be computed as 
$\rm M_Z(t)-M_d(t)$.

The  considerations discussed in section \ref{sec:sample} and, in
particular, the observed tight correlation between the observed $\rm M_{dust}$ and 
$\rm M_{H2}$ suggest that rapid and highly efficient grain growth takes place in
the dense molecular gas of high-redshift QSOs. We therefore
maximize the dust accretion process assuming that
all the gas-phase metals, $\rm M_{Z}^{\rm diff}(t)-M_{d}^{\rm diff}(t)$, that collapse in 
MC (during the clouds condensation stage) and survive the SF process 
(are not reincorporated into stars) can be accreted onto dust grains. Hence, 
in each galaxy of the merger tree the grain growth rate is computed as: 
\begin{eqnarray}
\rm 
    M_d^{MC}/\tau_{acc} & = & \rm [Z_{diff}(t)-D_{diff}(t)][\dot{M}_{cond}(t) - SFR(t)] \nonumber \\
     & & \rm = [Z_{diff}(t)-D_{diff}(t)](\alpha_{MC}-1)SFR(t)
\end{eqnarray}
\noindent
where $\rm Z_{diff}(t)-D_{diff}(t)$ is the mass fraction of gas-phase metals in the 
diffuse phase. Hence, in our formulation the dust accretion timescale depends on local 
conditions (metallicity and gas mass in MC), and shows a large variation among different
progenitor galaxies, with average values ranging between a few Myr to a few tens of Myr.

\section{Model results}
\label{sec:results}

All the quasars in the $5\leq z \leq 6.4$ sample show similar properties 
in terms of the SMBH, molecular gas, dynamical and dust masses. 
Thus, one can expect that their evolution has occurred along 
similar pathways. To investigate this issue, we apply the 2-phase fiducial 
model to 5 out of the 12 quasars listed in Table~\ref{tab:sample}: \\
\noindent
J1148 at $z=6.4$, J2310 at $z=6$, J0927 and J1044 at $z=5.8$, and J0338 at 
$z=5$. 
As shown in Fig.~\ref{fig:bhrel}, J1148 and J1044
are the most peculiar objects in the sample, presenting the largest 
deviation from the  local $\rm M_{BH}-M_{star}$ relation. Conversely, J0927 and 
J0338 are located closer to the local value in the BH-dynamical/stellar mass
plane, but still outside the observed scatter. Finally, J2310 is the brightest
quasar in the sample, with the largest estimated dust mass. 

For each quasar, we compute 50 different merger histories. 
In all the figures presented in this section, the solid lines indicate the 
results averaged over the 50 merger trees and the shaded area show the 
$1\sigma$ dispersion.
For all but the quasars J0338 and J1044, we use the 2-phase fiducial model
with the same parameters adopted in the model B3 presented in Valiante et al.
(2011) and the new free parameter $\rm \alpha_{MC}=3.08$ suited for J1148 
(see Table~\ref{tab:modelB3}).

\begin{table}
  \caption{Parameters adopted in the fiducial model (see text for details).}
  \label{tab:modelB3}
  \begin{tabular}{|c|c|c|c|c|c|c|}
    \hline
    {\bf QSO} & {\bf $\epsilon_{\rm quies}$} & {\bf $\sigma_{\rm burst}$} & {\bf $\epsilon_{\rm burst,max}$} & {\bf $\alpha$} & {\bf $\epsilon_{w,\textsc{agn}}$} & $\alpha_{MSc}$ \\
    \hline\   
    {\bf J0338} & 0.1   &  0.05   & 8.0  &   230   &  $5\times10^{-3}$ & 3.08\\
    {\bf J0927} & 0.1   &  0.05   & 8.0  &   200   &  $5\times10^{-3}$ & 3.08\\ 
    {\bf J1044} & 0.1   &  0.05   & 8.0  &   210   &  $2.5\times10^{-3}$ & 3.08\\
    {\bf J2310} & 0.1   &  0.05   & 8.0  &   200   &  $5\times10^{-3}$ & 3.08\\
    {\bf J1148} & 0.1   &  0.05   & 8.0  &   200   &  $5\times10^{-3}$ & 3.08\\
    \hline
  \end{tabular}
\end{table}
The QSO J0338 requires a higher BH accretion efficiency in the 
BHL formula (see eq.~6 in Valiante et al. 2011), 
$\alpha=230$, to compensate for the lower gas density at $z=5$.
Indeed, both in semi-analytical models and in hydrodynamical simulations, 
$\alpha$ parametrizes out poor knowledge of the gas density in the vicinity of 
the BH accretion radius. 
In the case of J1044, a higher BHL accretion efficiency, $\alpha=210$, combined 
with a lower BH-feedback efficiency, $\epsilon_{w,AGN}=2.5\times 10^{-3}$, are 
required in order to ensure the growth of the largest SMBH of the sample, with a
final BH mass of $\sim 10^{10}$ M$_\odot$ at $z=5.8$.  
Even if detailed local variations of the gas density can not be captured by a 
semi-analytical approach, in our model the average gas density, the rate of 
major mergers and the distribution of seed BHs in the first progenitors, which 
depend on redshift, all concur in shaping the evolution of the QSO.
For a discussion of the non-linear dependence of BH growth on the alpha 
parameter, the AGN wind efficiency, and the adopted mass of BH seeds we refer 
the interested reader to Valiante et al. (2011).

\begin{figure}
\includegraphics [width=8.8cm]{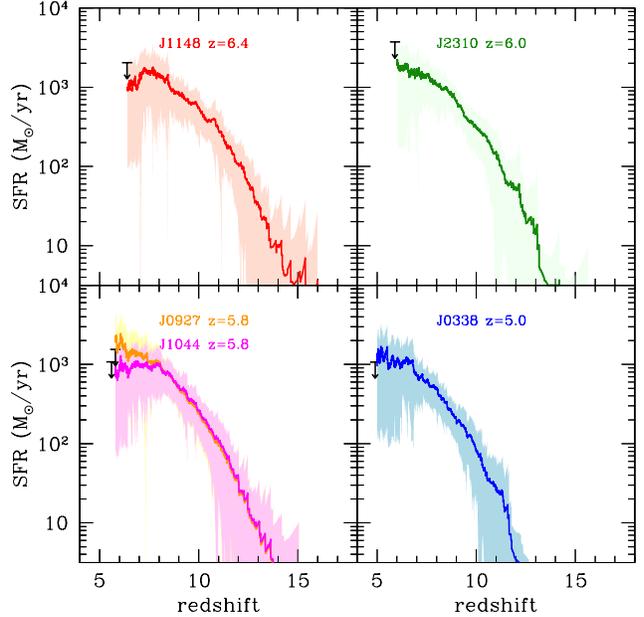}
\caption{Star formation rate as a function of redshift for 5 quasars:
 J1148 at $\rm z=6.4$ (upper left), J2310 at $\rm z=6$ (upper right), 
J0927 and J1044 at $\rm z=5.8$ (lower left) and J0338 at $\rm z=5$ 
(lower right). The solid lines indicate the average over 50 realizations of the merger 
  history of each object with the shaded regions representing the $1\sigma$ 
  dispersion. The data points in each panel indicate the IMF-corrected SFR
  inferred from the FIR luminosity (see text for details).} 
\label{fig:sfr} 
\end{figure}

In Fig.~\ref{fig:sfr} we show that all quasars follow a bursting SFHs with high final rates 
ranging between 800 and 2000 $\rm M_\odot/yr$.
These values are in good agreement with the rates computed from the FIR 
luminosity, once the correction due to the
(non-negligible) contribution of the AGN to dust heating is taken into account
(see discussion in Section \ref{sec:sample}). 
In all the panels, the data points indicate the SFRs corrected for the Larson IMF with
$\rm m_{ch} = 0.35 \, M_\odot$, which are approximately a factor 2 lower than the 
values for a Salpeter IMF quoted in Table~\ref{tab:sample}.
The down-turns and the modulations of the SFHs at 
$\rm z\leq 8$ are due to the negative effect of BH feedback, which drives a 
powerful outflow of material. 

\begin{figure}
\includegraphics [width=8.8cm]{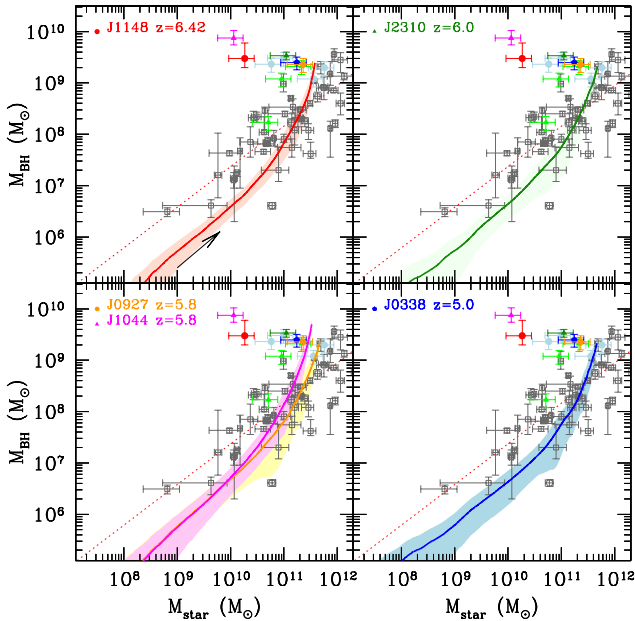}
\caption{Black Hole mass as a function of the stellar mass. 
         Solid lines are the redshift evolutionary paths (black arrow 
         in the upper lef panel) of the BH and stellar mass predicted by the 
         2-phase fiducial model}. Filled data points represent the high 
         redshift quasars, as labelled in each 
         panel, while open squares are a collection of local galaxies with the 
         empirical fit (dashed line) by Sani et al. (2011).
\label{fig:bhrel_models} 
\end{figure}

The corresponding evolutionary paths followed in the $\rm M_{BH} - M_{star}$ plane are
shown in Fig. \ref{fig:bhrel_models}; for reference, we also show the observed
values for high redshift QSOs (filled data) and local galaxies (open squares).
As expected, the final SMBHs are all reproduced by the model, while the high SF 
efficiencies result in final stellar masses in the range $(3-5)\times 10^{11}$ M$_\odot$, 
required to reproduce the observed dust mass and final SFRs. 
These are a factor of $3-30$ larger than the the observed values estimated as 
$\rm M_{dyn} - M_{H_2}$, which are shown by the filled data points in all panels. 
This tension is less critical for quasars like J0927 and J0338 which are found to 
lie within the scatter of the correlation observed for local galaxies.

The differences among different quasars can be explained by noting that, 
although similar final BH masses are produced and the same value for
the SF efficiency is adopted for all quasars (see Table~2),  the nuclear BHs grow at 
slightly different rates: on the one hand the assembly of the SMBHs of J2310, J0927,
and J0338 is slower than that of J1148, allowing a $\sim 1.5$ times larger 
final stellar masses; on the other hand, a SFH and a final stellar mass similar 
to that of J1148 are found for J1044, where the larger SMBH
mass requires a faster BH growth with respect, for example, to quasar 
J0927, observed at the same redshift. \\

\begin{figure}
\includegraphics [width=8.8cm]{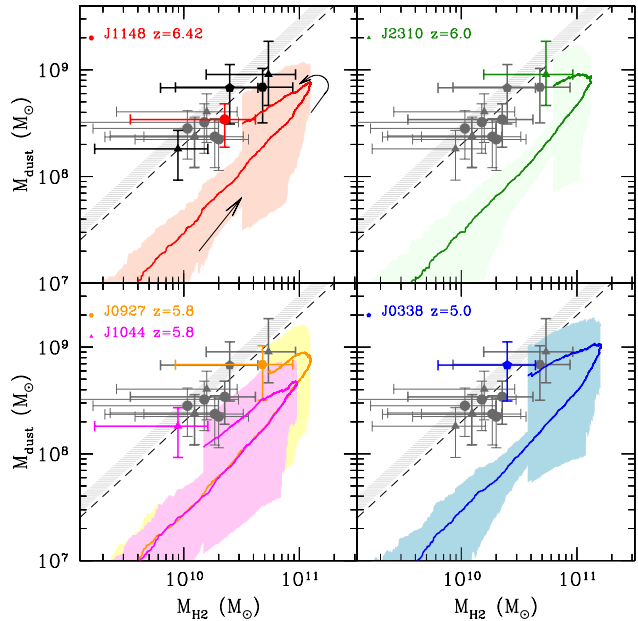}
\caption{Dust mass as a function of the molecular gas mass. This figure shows the 
         redshift evolution predicted by the 2-phase fiducial scenario (solid lines).
         The evolutionary path, in redshift, is traced by the black 
           arrows in the upper left panel. 
         The data points, line and shaded area are the same as in Fig.
         \ref{fig:dustMh2}. In all panels the labels indicate the modelled quasar.} 
\label{fig:dustMh2_models} 
\end{figure}

\begin{figure}
\includegraphics [width=8.8cm]{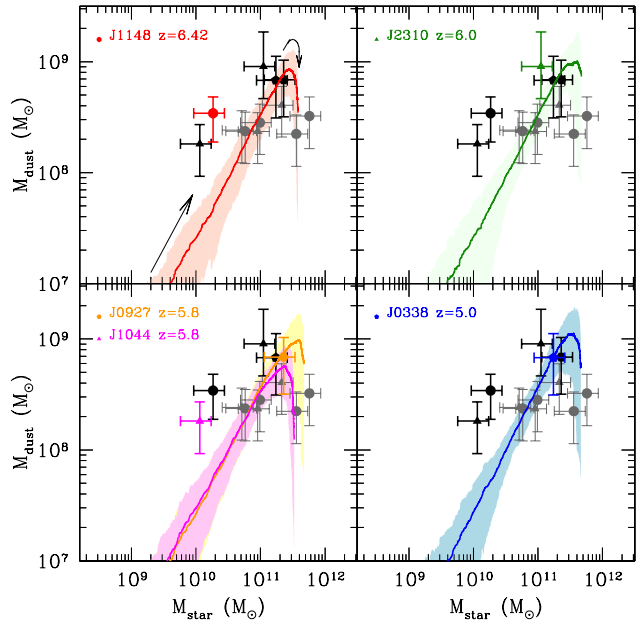}
\caption{Dust mass as a function of the stellar mass as predicted by the 
  2-phase fiducial scenario (solid lines). 
  The black arrows in the upper left panel indicate the redshift evolution of 
  the two quantities. The selected quasar and the corresponding model are 
  labelled in all panels. 
  Data points are the same as in Fig.~\ref{fig:dustMstar}}.
\label{fig:dustMs_models} 
\end{figure}

Finally, Figs.~\ref{fig:dustMh2_models} and \ref{fig:dustMs_models} show the 
evolution of the mass of dust as a function of the molecular gas and stellar 
masses, respectively.  

The model well reproduces the observed dust and molecular gas masses but 
over-predicts the final stellar mass. The curves show the redshift evolution 
of these quantities as predicted by the fiducial model B3 for each selected 
QSO. Following the hierarchical build-up of the host galaxy, the dust mass 
increases with increasing molecular and stellar masses (see black arrows in 
the upper left panels), driven by the enrichment of the ISM with heavy 
elements and dust and by efficient grain growth in MCs. At the final redshift, 
each evolutionary track matches the corresponding observed data point in the 
$\rm M_{dust} - M_{H_2}$ plane, but it is off-set toward larger stellar masses 
in the $\rm M_{dust}- M_{star}$ plane. 
These evolutionary paths confirm for a larger sample of QSOs the results 
obtained from our previous analysis that was applied to QSO J1148 only 
(Valiante et al. 2011, 2012; Schneider et al. 2014b).

In the picture of the Galaxy-BH co-evolution, the so-called 
\textit{cosmic cycle} (Hopkins et al. 2006), the onset of the active 
quasar phase follows in 
time a stage in which the emission of the central source is completely 
obscured by the surrounding optically thick material. 
At this stage, the BH continues to grow in the buried quasar while a 
strong starburst ($\geq 1000$ M$_\odot$/yr) is ongoing. 
These are the properties of Sub Millimeters Galaxies (SMG) that are observed 
to have dust and molecular/stellar masses comparable to high redshift quasars 
hosts (Santini et al 2010; Michalovski et al. 2010; Magnelli et al. 2012).

In the models shown in Figs.~\ref{fig:dustMh2_models} and \ref{fig:dustMs_models}, 
the transition between a starburst-dominated and a QSO-dominated
evolution is regulated by the growth of the central SMBH. 
When the stellar bulge reaches a mass $\rm M_{star} = (2 - 4) \times 10^{11} M_{\odot}$
the ISM is enriched by a large dust mass, $\rm M_{dust} = (0.5 - 1) \times 10^9 M_{\odot}$,
with a dust-to-gas ratio $\rm D \approx 1/200$, comparable to the values inferred for SMGs
and ULIRGs (Santini et al. 2010). At this stage, the nuclear BH has already grown to a mass 
$\sim 2\times10^8 - 10^9$ M$_\odot$, a strong energy-driven wind starts to clear up the ISM of dust and gas through
a large outflow, damping the star formation rate and un-obscuring the line of 
sight toward the QSO (see the hook-like shapes in the tracks of Figs.~\ref{fig:dustMh2_models} and \ref{fig:dustMs_models}).
For the QSOs that we have investigated, we predict a final gas 
(molecular+atomic) mass of $\sim (2-8)\times 10^{10}$ M$_\odot$ and an 
AGN-driven gas outflow rate ranging between $(4-6)\times 10^{3}$ M$_\odot/$yr. 
At these large rates, AGN-driven winds would be able to completely deplete the 
host galaxies of their gas content in less than $\sim 20$ Myr, shutting down 
both the star formation and the BH activity, and leaving behind a dead quasar
that will presumably evolve in a red (cD) galaxy. In other words, our model 
predicts the active quasar phase to last $\sim 10^7$ yr. This value is in 
perfect agreement with typical quasar lifetime values ($10^6-10^8$ yr) required
to match the present-day BH mass function and the QSO luminosity function at
$z=3$ (Haiman \& Loeb 1998; Martini et al. 2004), as well as with estimates 
obtained through the transverse proximity effect (Worseck 2007; Gallerani 2008).

\section{The SMG progenitors of high-z QSOs hosts}
\label{sec:smg}

All the evolutionary tracks discussed in the previous Section represent the 
time evolution of the global SFH, dust, stellar, and gas content of the host 
galaxy, without discriminating among the properties of single progenitor 
systems of the final QSO host. In reality, at each redshift a given QSO host 
is characterized by a large number of progenitor galaxies, each one 
characterized by its specific evolution. Hence, the hypothesis that 
high redshift QSOs have passed through an SMG-phase has to be tested against
the properties of their individual progenitors. 

We focus on two QSOs, J1148 and J0338, that 
represent the most and the less distant QSOs among the objects that we have 
considered in the present study.
Following Hayward et al. (2013), we compute the flux density at 
$\rm 850 \mu m$ of each progenitor galaxy in the merger trees as:
\begin{equation}
\rm 
  S_{850\mu m} = 0.81 \, \rmn mJy \, \left ( \frac{SFR}{100 \, \rmn M_\odot yr^{-1}}\right )^{0.43} 
                \left ( \frac{M_{d}}{10^8 \rmn M_\odot}\right )^{0.54}. 
\label{eq:flux}
\end{equation}
\noindent
Among all the progenitors, we classify as starbursts those which are 
characterized by a SFR$>100$ M$_\odot/$yr, and as SMG starbursts which also 
have a dust mass $\rm M_{d}>10^8$ M$_\odot$ and a flux density of 
$S_{850\mu m}>3$ mJy (Coppin et al. 2008; Michalovski et al. 2010, 2012; 
Magnelli et al. 2010,2012; Hayward et al. 2011, 2012).

\begin{figure}
\includegraphics [width=8.8cm]{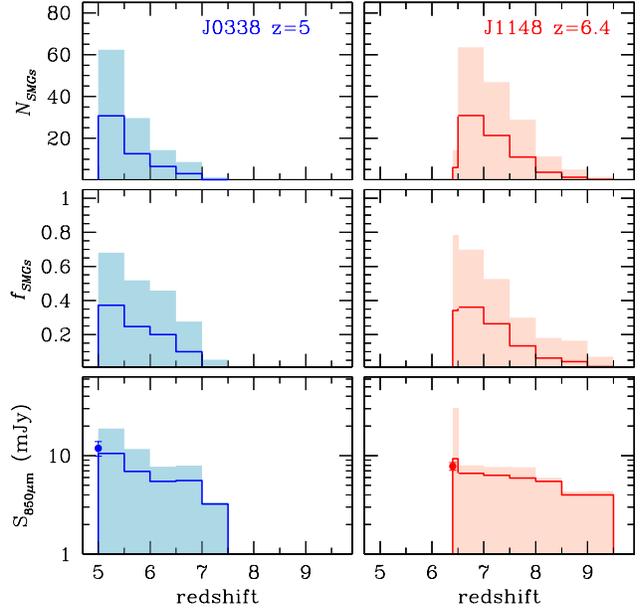}
\caption{Number, $N_{SMGs}$ (upper panels), fraction, $f_{SMGs}$ (middle pane) 
         and 850$\mu$m flux density, $S_{850\mu m}$ (lower panels) as a 
         function of redshift of SMG precursors of J1148 (left panels) and
         J0338 (right panels).  
         Solid lines are the averages over 50 different merger tree 
         realizations of each quasar with shaded regions representing the 
         1$\sigma$ error.
         Data points in the lower panels indicate the two QSOs observed flux densities (Carilli \& Walter 2013 and references 
         therein).} 
\label{fig:SMGs} 
\end{figure}

In Fig.~\ref{fig:SMGs} we show the redshift distribution of the average number 
of SMG, \textit{N}$_{SMGs}$, and of the average fraction of starbursts which
are SMG, $f_{SMG}$; for both quasars, each quantity is averaged over 50 
different merger trees.
As it can be seen in the upper and middle panels, the first progenitor 
galaxies of J1148 (J0338) start to enter the SMG phase at redshifts 
$\rm z \approx 9.5 \, (7.5)$; hence in the last $\approx 0.4-0.5$~Gyrs
of their evolution. 
At earliest epochs, there are many starbursting 
progenitors but the dust content in their ISM is too low to power a significant
FIR luminosity.  Thereafter, an increasing number of progenitors are 
classified as SMGs, reaching up to $\sim 40\%$ of the total starbursts 
progenitors around the final redshift $\rm z \sim 6.4~(5)$. 
As it is shown in the bottom panels of the figure, both QSO hosts meet the criteria 
to be characterized as SMGs.

We conclude that if the semi-empirical formula (eq.~\ref{eq:flux}) 
that we have used to compute the flux density at 850 $\mu$m can be 
extrapolated at redshifts $\rm z > 7$, we should expect to observe
SMG precursors of high-z QSOs up to $\rm z \sim 7 - 8$ although their number 
rapidly decreases with redshift.

Current surveys show that observed redshift distribution of SMGs has a maximum 
at $z\sim 2-3$ (Chapman et al. 2005; Yun et al. 2012, Smolcic et al. 2012).
The redshift distribution and evolution of SMGs appear to be very similar
to those of QSOs, suggesting a link between the two populations of objects 
(Maiolino 2008).
The exact SMGs number counts at $z>4$ require the identification of the optical
(or near-IR) counterparts of SMGs to determine their redshift through 
spectroscopic followup. Most of the SMG detections have been obtained so far
through single dish telescopes, whose angular resolution is low 
($11^{''}-18^{''}$). This implies that several optical/near-IR candidate 
couterparts are found within the telescope beam. It is therefore necessary to 
obtain mm-submm observations of high-z SMGs at higher angular resolution to 
better constrain their redshift evolution.
PdBI observation, characterized by angular reolution $\sim 1.5^{''}$, have
indeed shown that the surface density of $z>4$ SMGs is higher than predicted
by models (Smolcic et al. 2012 and references therein).
Moreover, submm color-selection techniques have recently enabled the
detection of a massive starburst galaxy at redshift 6.34 (Riechers et al. 
2013). Although these observations are very challenging, larger and 
deeper mm surveys in the future will allow to better constrain the evolution 
of starburst galaxies at high redshift.

\section{Discussion and Conclusions}
\label{sec:disc}

In this paper we have presented a 2-phase semi-analytical model for the 
formation and evolution of high redshift quasars and their host galaxies in the 
framework of hierarchical structure formation.
The model allows to investigate possible pathways to the assembly of the 
first SMBHs and of their host galaxies through cosmic times. 
The evolution of the mass of gas, metals and dust are consistently followed in 
both the diffuse and dense ISM as described in Section~\ref{sec:gamete2p}.
This model has been applied to a sample of quasars observed at 
redshifts $z=5$ and $z=6.4$ that show similar observed properties.

To explain the observed properties of these quasars, our study points to a 
common evolutionary scenario: during the hierarchical assembly of the host DM 
halo, stars form according to a standard IMF (hence a Larson IMF with a 
characteristic mass of $m_{ch} = 0.35$ M$_\odot$), via quiescent 
star formation and efficient merger-driven bursts. 
At the same time, the central BH grows via gas
accretion and mergers with other BHs. As the BH reaches a threshold mass of
$\sim 2\times 10^8 - 10^9$ M$_\odot$ 
its growth becomes more rapid and the predicted $\rm M_{BH}-M_{star}$ evolution 
steepens (see fig.~\ref{fig:bhrel_models}).
In this scenario, all the QSOs host galaxies are characterized by final 
stellar masses in the range $(3-5)\times 10^{11}$ M$_{\odot}$, a factor 3-30 
larger than the maximum values allowed by the observed $\rm M_{dyn} - M_{H_2}$.
Note that similar conclusions have been found  by 
numerical simulations aimed to describe both the formation mechanism, metal 
enrichment and dust properties of $z\sim 6$ QSOs, among which J1148, 
(Li et al. 2007, 2008), which also predict by a final stellar mass of 
$\sim 10^{12}$ M$_\odot$.

To better understand this apparent tension among model predictions and 
observed data, in our 
previous investigations, which have focused on J1148, we have explored 
alternative scenarios that we critically propose here for discussion: \\

\noindent
{\bf A top-heavy stellar initial mass function?} \\
A top-heavy IMF represents an alternative way to increase the integrated dust 
and metal yields without requiring a higher star formation efficiency and thus 
a larger final stellar mass. 
At high redshift, a stellar mass distribution biased towards more massive stars 
could be favoured by the physical properties of the ISM (Schneider \& Omukai 
2010; Smith et al. 2009; Jappsen et al. 2009; Klessen, Spaans \& Jappsen 2007).
In Valiante et al. (2011), we have shown that models with Larson
IMF and a characteristic mass $m_{ch}=5.0$ M$_\odot$ (top-heavy IMF)
can reproduce the chemical properties of the host galaxy of J1148 without
exceeding the upper limit set by the observed $\rm M_{dyn} - M_{H_2}$ on the 
final stellar mass. However, this requires a lower SF efficiency and thus a 
lower SFR, which at $\rm z = 6.4$ is $66\pm 59$ M$_\odot/$ yr\footnote{this 
value is the average over 50 different merger tree realizations with the 
error representing the $1\sigma$ dispersion}. 
This is too small to account for the observed FIR luminosity of the QSO of 
$2.2\times 10^{13}$ L$_\odot$ (see Table~\ref{tab:sample}).
In fact, using the conversion factor between the SFR and the FIR luminosity for a 
top-heavy IMF (see Section~\ref{sec:FIRobs}), we find that $L_{FIR} = 1.9\times 10^{12}$ 
L$_\odot$, a factor of $\sim 10$ smaller than the observed value. 
This simple argument has been further confirmed by a detailed radiative transfer model 
(Schneider et al. 2014b). 
Hence, a top-heavy IMF model could accomodate the tension beween the dust and stellar 
masses but at the price of underpredicting the final SFR and FIR luminosity. \\

\noindent
{\bf A larger stellar dust yield?}

\noindent
An alternative solution could be to assume more efficient sources
of dust, hence to increase the stellar dust yield.
In our model, dust is produced by the two 
main stellar sources, AGB stars and SNe (Valiante et al. 2009, 2011). 
Depending on the stellar progenitor mass and initial metallicity,
AGB stars with masses in the range $[1-7]$ M$_\odot$ can 
release $10^{-3}-10^{-2}$ M$_\odot$ of dust (Ferrarotti \& Gail 2006;
Zhukovska et al. 2008). Dust formation in SN ejecta represents a rapid and efficient
way to enrich the ISM. SN dust yields for stars in the mass range
$[8-40]$ M$_\odot$ with metallicities $\rm 0 \le Z \le 1\,Z_{\odot}$ have been
taken from the grid developed by Bianchi \& Schneider (2007). These authors
find that $\rm (0.1 - 0.6)~M_{\odot}$ of dust form in the ejecta but that 
only between 2 and 20 per cent of the initial dust mass survives the passage of the reverse shock, 
on time-scales of about $4-8 \times 10^4$~yr from the stellar explosion,
depending on the density of the surrounding interstellar medium. Our fiducial model
assumes moderate destruction by the reverse shock, with effective SN yields of $10^{-2}-10^{-1}$ M$_\odot$
(Valiante et al. 2009).
Assuming these dust yields, it has been shown that the dust mass released by stellar
sources only (SNe and AGB stars) is not enough to explain the dust mass observed in
high redshift QSOs (see fig.~\ref{fig:dustMstar}) and grain growth in MC
has been invoked or adopted in models as a possible solution 
(e.g. Draine 2009; Michalowski et al. 2010, Valiante et al. 2011; Mattsson et 
al. 2011; but see also Zafar \& Watson 2013). 
Similar conclusions have been recently drawn by Rowlands et al. 
(2014) for SMG. They find that the dust mass observed in a 
sample of high-redshift ($\rm z > 1$) SMG requires much higher 
SN yields and/or efficient grain growth in molecular clouds.

Recent {\it Herschel} observations have shown that previous
detection of dust in SNe and SNRs based on mid-IR photometry
may have missed the dominant cold dust components: indeed, 
a dust mass of $(0.4-0.7)$ M$_\odot$ has been detected in SN1987A
(Matsuura et al.  2011) and comparable values have been observed in
Cas A (Barlow et al. 2010; Nozawa et al. 2010) and the Crab (Gomez et al. 2012).
While this is certainly an important confirmation of theoretical models, 
none of the above SNR is old enough (ages $< 10^3$~yr) 
for the reverse shock to have significantly affected the newly formed dust. 
In addition, even adopting maximally efficient SN yields, 
the mass of dust in high-z objects could not have originated by
stellar sources only, unless dust destruction by interstellar shocks is 
neglected (Dwek. Galliano \& Jones 2007; Gall et al. 2011; Zafar \& Watsson 2013). 
While grain destruction in the ISM is still subject to many uncertainties, 
theoretical models show that sputtering in gas-grain collisions and vaporization in 
grain-grain collisions can be very efficient (Jones et al. 1996; 
Jones \& Nuth 2011, Jones 2012; Bocchio et al. 2012; Jones et al. 2013; 
Asano et al. 2013).
Silicate and carbon dust destruction occurs on timescales $\sim (200 - 400)$~Myr that 
are comparable or even shorter than the evolutionary timescales of high-z galaxies.  
Observations indicate that dust destruction takes place in
regions of the ISM shocked to velocities of the order of 50 - 150 km s$^{-1}$ 
(Welty et al. 2002; Podio et al. 2006; Slavin 2009). 
In addition massive gas outflows on galactic scales have been observed for 
starbursts and QSOs both in the local Universe and at high-z (Feruglio et al. 
2010, Nesvabda et al 2010, 2011; Maiolino et al. 2012; Cicone et al. 2012). 
An outflow rate of $> 3500$ M$_\odot/$ yr has been inferred from CII 
observations of J1148 (Maiolino et al. 2012), in good agreement with model 
predictions (Valiante et al. 2012). 
In these extreme environments, it is hard to believe that all the newly formed 
dust injected by stellar sources is conserved in the ISM, without being 
destroyed or being ejected out of the galaxy. Hence we conclude that maximally 
efficient stellar dust yields may provide a solution only if all the stellar 
dust injected in the ISM is conserved, without destruction and/or ejection. \\

\noindent  
{\bf Shorter evolutionary timescale?}
\noindent
The fiducial scenario that we have presented predicts that high redshift
QSO hosts  are characterized by SFR $\sim (1 - 2 )\times 10^3$~M$_\odot/$yr
in the last $200-300$~Myr of their hierarchical assembly. The associated
stellar mass formed in this starburst is $\rm M_{star} = (2 - 3)\times 10^{11} M_{\odot}$
and exceeds the upper limits inferred from $\rm M_{dyn} - M_{H_2}$. 
For the two QSOs J1044 and J1148, the stellar mass is within the observed upper limits
only assuming a burst of shorter duration, $< 10-20$~Myr.  This time-scale
is comparable to the lifetime of a $10 - 20 \, \rm M_{\odot}$ star
and can be shown to be too short to explain the observed dust masses
of QSO hosts and the mass of the nuclear SMBHs.

For a 20 Myr burst, the IMF-weighted stellar dust yield
ranges between
$8.7 \times 10^{-5}$ ($1.2 \times 10^{-3}$) and $2.4 \times 10^{-4}$ ($3.3 \times 10^{-3}$)
for a standard and a top-heavy IMF, and the values in parenthesis indicate the corresponding
yields when maximally efficient SN dust models with no reverse shock destruction are considered
(Valiante \& Schneider 2014). Hence, the dust mass produced by 
a young starburst is at most $7 \times 10^7$ M$_\odot$, too small to account for the observed
dust masses.

Moreover, if QSO host galaxies were only 20 Myr old, their SMBH should have grown from a seed of
comparable mass, $(1-6)\times 10^9$ M$_\odot$, assuming continuous 
accretion at the Eddington rate. This is about $3-4$ orders of magnitude higher that the 
heaviest seed BHs expected to form by gas- or stellar-dynamic direct collapse (Bromm \& 
Loeb 2003; Begelman, Volonteri \& Rees 2006; Lodato \& Natarajan 2007; Omukai 
et al. 2008, Devecchi \& Volonteri 2009; Bellovary et al. 2011).

Hence, we conclude that observations do not support the idea that high-z QSO hosts
have evolved to their final $\rm M_{BH} - M_{star}$ in only 20 Myr. \\

\noindent
{\bf Are dynamical mass measurements missing some of the stars?}

\noindent
Possible solutions to the discrepancy between the 
stellar mass predicted by theoretical models and the upper limits
derived from  $\rm M_{dyn} - M_{H_2}$ 
can be found in the assumptions made to estimate the dynamical and
molecular gas masses from the observations. 
As we have discussed in Section 2, the largest uncertainties are 
due to the assumed properties of the CO-emitting disk; in particular, the
radius $\rm R$ and the inclination angle $i$. 

\begin{figure}
\includegraphics [width=8.8cm]{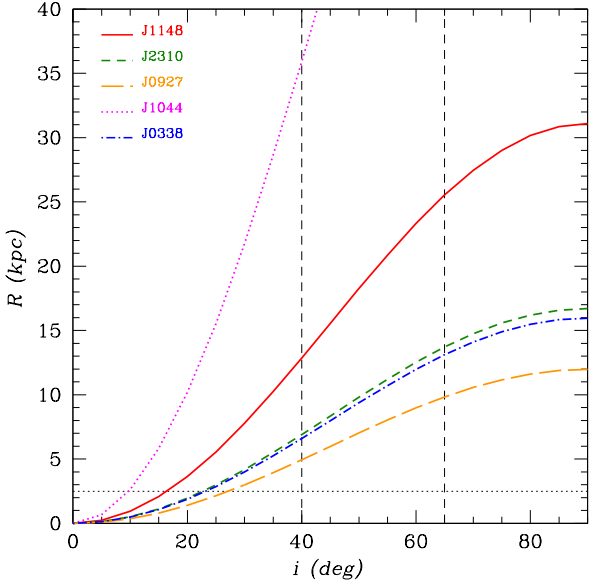}
\caption{The radius of the disk enclosing the dynamical mass
predicted by the fiducial models as a function of the inclination angle.
Each line corresponds to a give QSO, as labelled in the figure. The two
vertical dashed lines correspond to the inclination angles of $65^o$ and
$40^o$ adopted in the literature for high-z QSOs and used in Section 2.
Similarly, the horizontal dotted line shows the 2.5 kpc value.}
\label{fig:radangle} 
\end{figure} 

Observational constraints on these two parameters are available only
for J1148, in which high resolution imaging has enabled to resolve CO emission
within $\rm R=2.5$ kpc and $i=65^o$ (Walter et al. 2004, Riechers et al. 2009). 
As we have discussed in Section~2.2, a disk radius of $\rm 
R=2.5 kpc$ and inclination angles $i=40^o,65^o$ have been commonly assumed to
describe high-z QSOs (Wang et al. 2010, 2013). 
It is important to note, however, that such compact regions represent 
only a very small fraction of the virial radius,  $\sim 1/40$, for these
very massive halos. This may be an indication of a more extended 
distribution of their actual baryonic content (Khandai et al. 2012, 
see discussion below).

To accommodate the stellar masses predicted by the fiducial models,
$M^{model}_{star}$, the 
minimum upward correction to $\rm M_{dyn}$ ranges between 2 - 3
for J2310, J0927, and J0338; larger corrections are required for 
J1148 (a factor 10) and J1044 (a factor 15). 
In Fig.~\ref{fig:radangle}, for each QSO we show the properties of the disk 
that would yield, for a given CO line FWHM, the minimum dynamical mass 
expected by the models, $M^{model}_{dyn}$: 
$\rm R= M^{model}_{dyn} G/v^2_{circ}$, where $\rm v_{circ}=(3/4)FWHM_{CO}/$sin $i$,
$\rm M^{model}_{dyn}\geq M^{model}_{star}+M^{obs}_{H2}$ and $M^{obs}_{H2}$ is the
observed $\rm H_2$ mass given in Table~\ref{tab:sample}.
It is clear from the figure that a given dynamical mass can be accommodated
within a radius that grows with the inclination angle.

For J1148, the predicted dynamical mass $M^{model}_{dyn}\sim 4.23\times 10^{11}$ 
M$_\odot$ would be enclosed within a radius of $\sim 25$ kpc or within a highly
inclined disk $i < 15^o$ (Wang et al. 2010).
However, high-resolution  Very Large Array (VLA) 
observations of the CO emission show that the emission breaks up into two 
regions, separated by 1.7 kpc, possibly revealing an ongoing merger (Walter et 
al.2004) and thus, indicating a more complex gas dynamics than described by a
simple disk model. 
 
Fig.~\ref{fig:radangle} shows that for QSOs 
which would require a relatively small correction to the 
inferred dynamical mass (J2310, J0927 and J0338), a disk radius in the range 
(5-8)~kpc would be adequate if seen with an inclination angle of $40^o$; 
conversely, a disk radius of 2.5~kpc would require an inclination angle of 
$i \sim 25^o - 30^o$. 
For J1044, in which the largest fraction of stars is missing, 
the predicted dynamical mass would be accommodated within  a radius of 
$\sim 35$~kpc or $i=10^o$.

It is important to note that recent ALMA observations have marginally
resolved the [CII] emission in QSOs J2310 and J1044 (Wang et al. 2013).
For J2310, the estimated inclination angle is $i = 46^o$, larger than the 
value used to infer the dynamical mass from the CO. However, Wang et al. (2013)
underline that the measurement of the sizes of high redshift sources and thus 
of the disk inclination angles are still highly uncertain even at the 
$\sim 0.7^{''}$ spatial resolution with ALMA.
For J1044, ALMA observations pointed out that there are differences in the 
[CII] and CO spectra which may indicate a more complex dynamics.
The [CII] line profile of this source has a larger redshift and a broader line 
width with respect to that of the CO(6-5) detection. These suggest either a 
difference in the kinematical properties of the two gas components or that a 
significant fraction of the CO emission may be undetected.

However, for most of these high redshift sources the apparent tension between
model predictions and observational data may be alleviated by modifying the 
adopted gas disk geometry or by allowing for more complex merger-like gas 
distributions. 
Note that Narayanan et al. (2009) have modeled CO molecular lines in 
high redshift SMGs via numerical simulations, showing that if SMGs are 
typically a transient phase of major mergers, the usage of standard CO 
techniques to infer physical quantities may lead to inaccurate measurements of 
the true enclosed dynamical mass by a factor $\sim 2$ from the actual value.
Deep imaging of the CO line emission with better measurements 
of the CO line profile and spatial distribution is needed to better constrain 
the dynamical masses of these systems.

The idea of a more complex and extended distribution of the stars in 
high-z QSO host galaxies, 
is supported by recent hydrodynamical simulations (Khandai et al. 2012).
The simulations show that quasar host galaxies at $z = 5$ are indeed compact 
gas-rich systems with the bulk of star formation occurring in the very inner 
regions. These regions are surrounded by a number of star forming clouds, 
providing a significant amount of stars, distributed on a larger scale, within 
the DM halo virial radius ($\sim 200/h$ kpc). 
This suggest that the regions in which the CO is observed (when spatially 
resolved) may not trace the spatial distribution of the stellar 
component of the whole galaxy.
\\

We conclude that high-z QSO host galaxies follow a complex evolution
and gain the bulk of their stellar mass content through intense
dust-enshrouded starbursts that occur as early as 500 Myr after the
Big Bang. The star formation history and metal enrichment of these
galaxies are tightly coupled to the growth of their nuclear black hole.
When the black hole has grown to a mass $\sim 2\times10^8 - 10^9$ M$_\odot$, 
a strong energy-driven wind starts to clear up the ISM of dust and gas through
a large outflow, damping the star formation rate and rendering the QSO
optically bright. At this stage, the stellar bulge has already grown to values
that exceed the upper limits inferred from dynamical mass and molecular gas
measurements. However, for most of these sources the apparent tension between
model predictions and observational data may be alleviated by modifying the
adopted gas disk geometry or by allowing for more complex merger-like
gas distributions. 
Deep imaging of the CO line emission with better 
measurements of the CO line profile and spatial distribution is needed
to better constrain the dynamical masses of these systems.

\section*{Acknowledgments}
We thank the anonymous referee for usefull suggestions and comments.
We thank E. Sani for providing useful data, X. Fan and P. Santini for fruitful 
discussion and comments.
SS acknowledges support from Netherlands Organization for Scientific 
Research (NWO), VENI grant 639.041.233.
The research leading to these results has received funding from the European 
Research Council under the European Union’s Seventh Framework Programme 
(FP/2007-2013) / ERC Grant Agreement n. 306476.
This research was supported in part by the National Science Foundation under 
Grant No. NSF PHY11-25915.


\label{lastpage}

\end{document}